\documentclass[9pt, conference]{IEEEtran}
\IEEEoverridecommandlockouts

\usepackage{cite}
\usepackage{amsmath,amssymb,amsfonts}
\usepackage{algorithmic}
\usepackage{graphicx}
\usepackage{textcomp}
\usepackage{xcolor}
\usepackage{caption}
\usepackage{multirow}
\def\BibTeX{{\rm B\kern-.05em{\sc i\kern-.025em b}\kern-.08em
    T\kern-.1667em\lower.7ex\hbox{E}\kern-.125emX}}
\begin{document}

\title{Evaluating the Impact of Discriminative and Generative E2E Speech Enhancement Models on Syllable Stress Preservation}

%
%
%

\author{\IEEEauthorblockN{Rangavajjala Sankara Bharadwaj, 
 Jhansi Mallela, 
Sai Harshitha Aluru, 
Chiranjeevi Yarra} 
\IEEEauthorblockA{Language Technology Research Centre, 
International Institute of Information Technology, Hyderabad, India \\ 
bharadwajsankara@gmail.com, 
jhansi.mallela@research.iiit.ac.in, 
chiranjeevi.yarra@iiit.ac.in}}

\maketitle
\begin{abstract}
Automatic syllable stress detection is a crucial component in Computer-Assisted Language Learning (CALL) systems for language learners. Current stress detection models are typically trained on clean speech, which may not be robust in real-world scenarios where background noise is prevalent. To address this, speech enhancement (SE) models, designed to enhance speech by removing noise, might be employed, but their impact on preserving syllable stress patterns is not well studied. This study examines how different SE models, representing discriminative and generative modeling approaches, affect syllable stress detection under noisy conditions. We assess these models by applying them to speech data with varying signal-to-noise ratios (SNRs) from 0 to 20 dB, and evaluating their effectiveness in maintaining stress patterns. Additionally, we explore different feature sets to determine which ones are most effective for capturing stress patterns amidst noise. To further understand the impact of SE models, a human-based perceptual study is conducted to compare the perceived stress patterns in SE-enhanced speech with those in clean speech, providing insights into how well these models preserve syllable stress as perceived by listeners. Experiments are performed on English speech data from non-native speakers of German and Italian. And the results reveal that the stress detection performance is robust with the generative SE models when heuristic features are used. Also, the observations from the perceptual study are consistent with the stress detection outcomes under all SE models.
\end{abstract}


\section{Introduction}
\vspace{-0.1cm}
\label{sec:intro}

Human-computer interaction (HCI) systems have become integral to everyday life, relying heavily on the clarity and precision of spoken language to function effectively. Among the various elements of spoken language, syllable stress plays a crucial role in shaping meaning, particularly in stress-timed languages like English. For instance, in the word \textit{permit} with the syllable sequence \textit{p ah - m ih t}, placing stress on the first syllable indicates a noun, while stressing the second syllable indicates a verb. Native English speakers naturally acquire the skill of placing stress, but non-native speakers often struggle with syllable stress due to the influence of their first language. To assist non-native speakers, Computer-Assisted Language Learning (CALL) systems \cite{levy1997computer} have been developed to automatically detect stress patterns in syllables and provide guidance. For these systems to be effective, the stress detection module must be highly robust to ensure accurate guidance \cite{ferrer2015classification}. However, since these modules are typically trained on clean, conditioned data, they can be prone to errors in noisy, real-world environments, which are often unavoidable.

To address the challenge of noise in real-world environments, two strategies can be employed: (1) training models with extensive, diverse noisy data to build robustness against noise, or (2) removing the noise from the speech before using for the downstream task. The first approach is often impractical due to its high computational cost, so in the literature, many systems prefer the second approach, known as speech enhancement. Speech enhancement (SE) is a process that improves speech quality and intelligibility by mitigating noise or reverberation \cite{benesty2006speech}. It is commonly used as a preprocessing step in applications such as automatic speech recognition (ASR), speech emotion recognition (SER), and automatic speaker verification (ASV) to enhance performance \cite{weninger2015speech, triantafyllopoulos2019towards, michelsanti2017conditional}. However, SE models face challenges related to latency, speech quality preservation, and model complexity. To address these, several approaches have been proposed in the literature, utilizing various modeling techniques each prioritizing different aspects while balancing trade-offs between computational efficiency, speech clarity, and real-time performance.

\begin{figure}
\vspace{1.05cm}
\centering
    \vspace{-1.2cm}
    \includegraphics[width=\columnwidth]{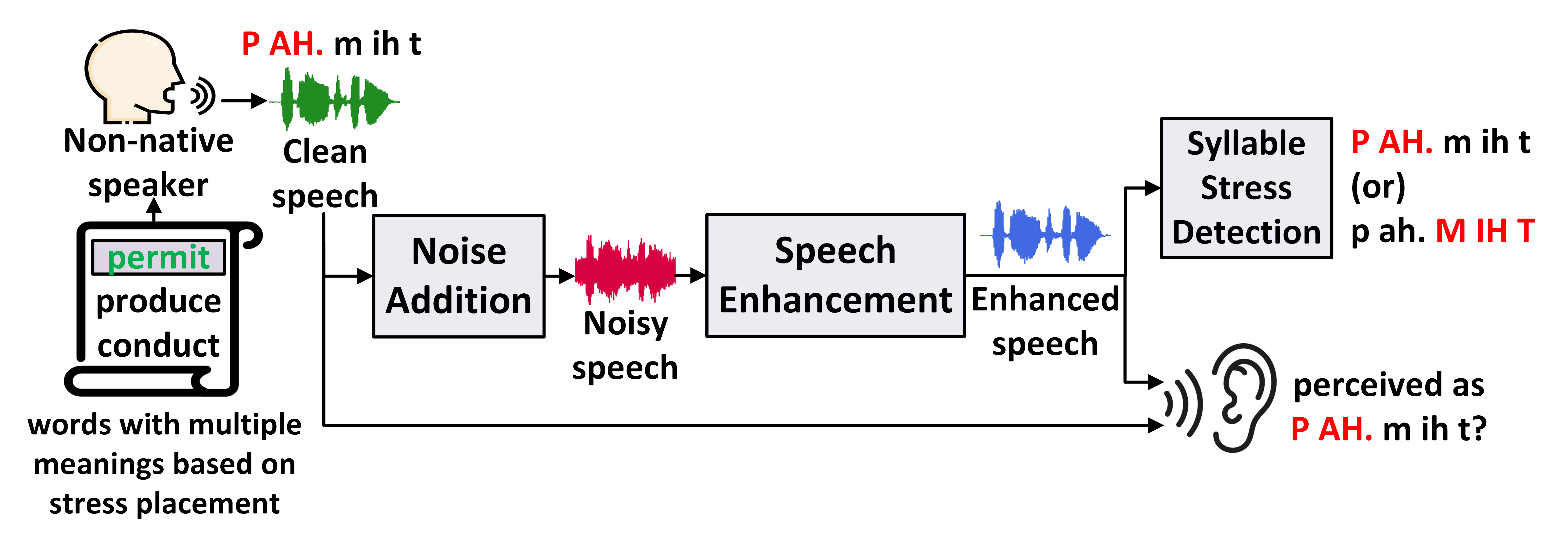}%
    \vspace{-0.25cm}
    \captionsetup{justification=centering}
    \caption{Overview of the proposed approach for evaluating the impact of speech enhancement in syllable stress preservation}
    \label{fig:Noisy_Enhanced}
    \vspace{-0.6cm}
    \end{figure}
 

Despite significant research on speech enhancement (SE) models in various speech-related tasks, their effect on syllable stress patterns has not been thoroughly examined. It remains uncertain how noise distorts these patterns and to what extent SE models are capable of preserving the original stress in speech. 
Syllable stress is a key aspect in English, where every word has one syllable that is more prominent or relatively emphasized than any other syllable in the same word. 
Automatic syllable stress detection involves classifying syllables into two categories (stressed/unstressed), making it a two-class classification problem. Several approaches for syllable stress detection have been proposed, primarily leveraging stress cues such as pitch, intensity, and duration \cite{delmonte1997slim, tepperman2005automatic}. Many machine learning (ML) models like boosting, bagging, decision trees, and support vector machines were used over the acoustic features computed on these stress cues for stress detection task \cite{johnson2015automatic}. An attention-based neural network combined with bidirectional LSTMs was used for stress detection, leveraging Mel frequency cepstral coefficients (MFCCs), energy, and pitch features \cite{xia2019attention}. Ruan et al. \cite{ruan2019end} utilized a transformer network for stress detection, while Mallela et al. \cite{mallela2023comparison} introduced a novel methodology with a joint optimization technique using variational autoencoders. All these studies have demonstrated the effectiveness of these methods under clean speech conditions. However, the challenge of accurately detecting stress patterns in noisy conditions and the effect of SE models in retaining stress cues is still not addressed.

The primary objectives of this study are to: 1) assess the performance of syllable stress detection under noisy conditions across various signal-to-noise ratios (SNRs), 2) evaluate the effectiveness of different state-of-the-art speech enhancement (SE) models in preserving stress patterns amidst noise and their impact on stress detection accuracy, and 3) conduct a perceptual study to determine whether automatic stress detection results align with human perception-based outcomes.
In this study, we address these objectives by exploring the effects of Gaussian noise at varying SNR levels using three state-of-the-art SE models, namely DTLN, Denoiser and CDiffuSE, belonging to either discriminative or generative modeling approaches with each addressing different challenges. We employ a state-of-the-art approach combining Variational Autoencoder and Deep Neural Networks (DNN) for automatic syllable stress detection. For experiments, we utilize the ISLE corpus, which includes polysyllabic English words spoken by non-native German and Italian speakers. We investigate two feature sets for stress detection: 1) heuristic-based acoustic and context features, and 2) self-supervised representations from wav2vec-2.0. 
From the experimental results, we observe that the stress detection performance is significantly degraded when self-supervised representations are used compared to heuristic-based features under 0 and 5 dB SNRs with respect to clean condition for both noisy and enhanced audios. 
Between discirminative and generative SE models, the later one consistently outperformed the former one under almost all SNR conditions.
Further, the results from the perceptual study align with the automatic syllable stress detection outcomes across all three SE models.

\section{Dataset}
\vspace{-0.15cm}
\label{sec:dataset}
For this study, we utilize two datasets from the ISLE corpus:
1) German: This dataset includes 3,733 speech recordings from 23 non-native English speakers of German origin, and 2) Italian: This dataset comprises 3,981 speech recordings from 23 non-native English speakers of Italian origin.

\textbf{Data Processing:}
Phoneme alignments for both datasets are first computed using an automatic force alignment process. These alignments are then meticulously reviewed and corrected by a team of five linguists to ensure accuracy. Next, syllable alignments are derived from the phoneme alignments using the P2TK syllabification tool.  The same group of linguists annotates syllable stress, ensuring that every word contains exactly one stressed syllable. Train and test sets are partitioned considering speaker nativity, age, gender, and language proficiency, to maintain a balanced distribution of stressed and unstressed syllables across both the sets \cite{menzel2000isle}. We consider only polysyllabic words with two or more syllables for the experiments following the work in \cite{yarra2017automatic, mallela2023comparison}. 

\begin{figure*}
\centering
    \vspace{-1cm}
    \includegraphics[width=0.85\textwidth]{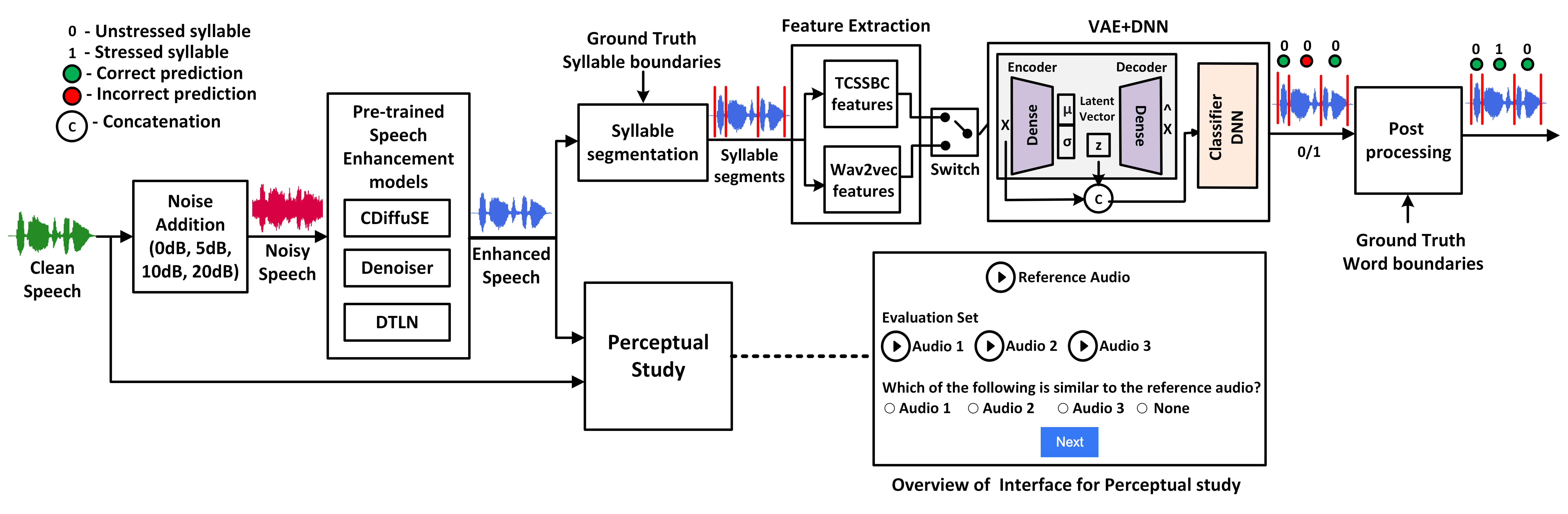}%
    \vspace{-0.3cm}
    \captionsetup{justification=centering}
    \caption{Block diagram demonstrating automatic syllable stress detection and perceptual study conducted on SE-enhanced speech. The dotted line connects to the overview of the interface used for the perceptual study.}
    \label{fig:Block_diagram}
     \vspace{-0.5cm}
    \end{figure*}

\section{Methodology}
\vspace{-0.15cm}
\label{sec:format}

In this section, we first describe the pipeline for the stress detection task and how it is structured to integrate various speech enhancement (SE) models. We then outline the specific SE models considered in this study. Finally, we explain the procedure for the perceptual study, which we conduct separately to explore the relationship between human perception outcomes and the automatic stress detection results.

\vspace{-0.1cm}
\subsection{Automatic syllable stress detection:}
\vspace{-0.1cm}
Figure \ref{fig:Block_diagram} illustrates the pipeline for the stress detection task. We start with clean speech signals and generate four sets of noisy data with the following signal-to-noise ratios (SNRs): 0 dB, 5 dB, 10 dB, and 20 dB, using white Gaussian noise. Each set is processed individually for the experiments. From these noisy speech signals, we obtain enhanced speech signals using three different SE models: CDiffuSE, Denoiser, and DTLN, which we discuss further.
Next, we perform syllable segmentation on each speech signal using syllable timestamps. For each syllable segment, we extract syllable-level features, which are then used to build a syllable stress detection classifier. As shown in the Figure \ref{fig:Block_diagram}, we consider two different set of features for the stress detection task, 1. Heuristic-based acoustic and context features and 2. Self-supervised representations (wav2vec 2.0). Finally, a postprocessing step ensures that each word contains only one stressed syllable.

\subsubsection{\textbf{Heuristic-based acoustic and context features}}
Syllable stress typically relies on three prominence measures: intensity, duration, and pitch, centered around the most sonorous sound unit in the syllable. In the literature, acoustic features were derived from contours like short-time energy, representing the intensity related measures. However, short-time energy contours can vary greatly across unstressed syllables and may introduce peaks due to surrounding sounds \cite{yarra2017automatic}. To address this, a sonority-based contour is proposed by combining sonority-motivated cues \cite{yarra2017automatic} with the short-time energy contour. Inspired by the work in \cite{yarra2017automatic}, we compute 19-dimensional statistical features, referred to as acoustic features (A). Further, in literature, it was stated that context significantly influences stress perception. In \cite{yarra2019comparison}, a 19-dimensional binary context feature vector is proposed for each syllable, capturing information about the syllable nucleus type, the preceding and following phoneme categories, and the word's position relative to pauses in the sentence. 

\subsubsection{\textbf{Self-supervised representations (wav2vec 2.0)}}
Recently, self-supervised models such as wav2vec 2.0 \cite{baevski2020wav2vec} and HuBERT \cite{hsu2021hubert} have gained considerable attention for their ability to learn robust speech representations without the need for manual labeling. Among these, wav2vec 2.0 is particularly notable for its effectiveness across a range of speech tasks. It excels in capturing spectral, phonetic, speaker, and semantic features from raw speech data \cite{sharma2022multi, wang2021fine}. Recently, it has been shown that wav2vec 2.0 features are able to capture the stress patterns as well \cite{mallela2024exploring}.
Following that, in this work, we leverage the 768-dimensional features extracted from wav2vec 2.0 for our experiments. Typically, wav2vec 2.0 features are obtained at the frame level. For our analysis, we aggregate these frame-level features to the syllable level by averaging them based on the syllable start and end timestamps. 

\vspace{-0.1cm}
\subsection{Speech Enhancement models (SE):}
\vspace{-0.1cm}
Most early works of SE are formulated in the time-frequency (spectrogram) domain and aim to approximate the clean spectrogram, from which they obtain the enhanced speech signal using the inverse short-time Fourier transform (iSTFT). However, the STFT is a generic transformation that may not be optimal for this task, and inaccuracies in phase reconstruction can limit the quality of the enhanced audio \cite{luo2019conv}. 
Adressing this, several methods have been developed that learn representations directly from raw speech signals using deep learning \cite{luo2018real, ronneberger2015u, luo2018tasnet}. While these approaches avoid some issues of spectrogram-based methods, they can still introduce unpleasant speech distortions and phonetic inaccuracies \cite{lu2022conditional}. In contrast, diffusion models, a category of generative modeling, progressively denoise and refine the speech signal through iterative steps. This method enhances speech quality and naturalness while effectively reducing noise, making diffusion models particularly promising for challenging enhancement scenarios.

In this study, we consider three state-of-the-art speech enhancement models for noise removal in the automatic syllable stress detection task, considering factors such as latency, model complexity, speech quality, and the type of modeling (discriminative vs. generative). These details are provided as follows:

\subsubsection{ \textbf{Dual-Signal Transformation LSTM Network (DTLN)}}
The Dual-Signal Transformation LSTM Network (DTLN), categorized as a discriminative model, is introduced in \cite{westhausen2020dual} for real-time speech enhancement (one frame in, one frame out). This model combines the short-time Fourier transform (STFT) with a learned analysis and synthesis basis in a stacked-network architecture, using fewer than one million parameters. With its low complexity and low latency due to the frame-by-frame processing, DTLN is suitable for real-time applications. Despite its simplicity, it is claimed that DTLN achieves high speech quality by efficiently extracting information from magnitude spectra and incorporating phase information from the learned feature basis. It is trained on 500 hours of noisy speech from the Librispeech corpus \cite{panayotov2015librispeech}, and the added  noise signals are from Audioset \cite{gemmeke2017audio}, Freesound, and DEMAND corpus \cite{thiemann2013diverse} with SNR levels ranging from -5 to 25 dB.


\subsubsection{ \textbf{DEMUCS-based real-time speech enhancement (Denoiser)}}
DEMUCS, which falls under the discriminative model category, is a novel architecture originally proposed for music source separation in \cite{defossez2019music}. To overcome the limitations of Conv-Tasnet \cite{luo2019conv}, which is the state-of-the-art approach for speech enhancement via noise removal, \cite{defossez2020real} proposed a real-time version of the DEMUCS architecture adapted for speech enhancement. It consists of a causal model based on convolutions and LSTMs, with a frame size of 40 ms and a stride of 16 ms. This results in moderate complexity but retains low latency for real-time applications. DEMUCS operates directly on waveforms, which enhances speech quality, especially when dealing with complex audio signals, due to its hierarchical generation and U-Net-like [21] skip connections. The model is trained for 400 epochs on the Valentini dataset \cite{valentini2017noisy} and for 250 epochs on the DNS dataset \cite{reddy2020interspeech}.


\subsubsection{\textbf{Conditional diffusion probabilistic model (CDiffuSE)}}
The Conditional Diffusion Probabilistic Model (CDiffuSE), a generative model, leverages diffusion probabilistic methods for speech enhancement \cite{sohl2015deep}. It converts clean data into an isotropic Gaussian distribution through a step-by-step process and then restores the clean input by removing noise iteratively. While vanilla diffusion models assume isotropic Gaussian noise, CDiffuSE, as proposed by \cite{lu2022conditional}, adapts to non-Gaussian noise characteristics for better real-world performance. While this approach provides high speech quality, it comes at the cost of high complexity and higher latency due to the iterative process required to remove noise. CDiffuSE model is trained on on the VoiceBank-DEMAND dataset \cite{valentini2016investigating} which consists of 30 speakers from the VoiceBank corpus \cite{veaux2013voice}. And, it is shown that CDiffuSE maintains strong performance when regression-based approaches such as Demucs \cite{defossez2020real} and Conv-TasNet \cite{luo2019conv} collapse.

\vspace{-0.2cm}
\subsection{Perceptual study:}
\vspace{-0.1cm}
In English, many words belong to two grammatical categories depending on the placement of syllable stress. For instance, words like `\textit{permit, produce}' can function as either a noun or a verb based on whether the stress falls on the first or second syllable. To evaluate how well SE-enhanced audio preserves syllable stress patterns, we conduct a perceptual study with 25 subjects, aged 20-25, fluent in English. Subjects assess 50 audio samples (25-GER and 25-ITA), each representing a word that can belong to two grammatical categories depending on the placement of stress.
This study involves the subjects to identify the best match between clean and the respective three enhanced audios in terms of stress placement by asking them to listen to those audios.
The main objectives of this perceptual study are: 
1) To identify which SE model most effectively retains stress patterns in enhanced audio, and 2) To investigate if the SE model that achieves the highest accuracy in automatic stress detection also produces enhanced audio that listeners find most similar to the clean reference. This comparison helps evaluate whether SE models that excel in automatic detection also effectively preserve stress patterns in a way that aligns with human perception.
\section{Experimental setup}
\label{sec:pagestyle}

For experiments of automatic syllable stress detection, following the work in \cite{mallela2024exploring}, we consider both heuristics-based acoustic and context features (38-dimension) and self-supervised wav2vec 2.0 representations (768-dimension).
In this study, we implement the state-of-the-art approach that jointly optimize VAE and DNN for syllable stress detection as proposed in \cite{mallela2023comparison}. This method leverages the VAE’s capability to learn comprehensive data distributions and generate meaningful latent representations, while the DNN focuses on extracting task-specific implicit representations.
We use a 5-fold cross-validation approach, where the training set is split into five equally sized groups with balanced stressed and unstressed syllables. In each fold, four groups are used for training and one for validation, rotating in a round-robin fashion. Both training and testing sets are normalized using the mean and standard deviation from the training set, and performance is evaluated based on the average classification accuracy across the five folds. \textbf{Architecture details:} VAE consists of 1 hidden layer each
in encoder and decoder with Relu activation function.  DNN model is built with
5 layers [hidden units: 64, 32, 16, 4, 1]. All the DNN and VAE parameters like number of layers, and number of nodes in each layer are optimal and we choose them by
maximizing the performance on the validation set.

\section{Results and Discussion}
\vspace{0.25cm}
\label{sec:typestyle}

\begin{table}[]
\scriptsize
\centering
\captionsetup{justification=centering}
\caption{\footnotesize{Classification accuracies with and without (in brackets) postprocessing, using acoustic and context features, for speech processed by different SE models under various SNR conditions. Accuracies for GER and ITA clean data are 93.36 (90.42) and 92.54 (89.36) respectively}}.
\begin{tabular}{|c|l|cccc|}
\hline
\multicolumn{1}{|l|}{\multirow{2}{*}{\textbf{\begin{tabular}[c]{@{}l@{}}Acoustic + \\ Context\end{tabular}}}} &
  \multirow{2}{*}{\textbf{Condition}} &
  \multicolumn{4}{c|}{\textbf{SNR}} \\ \cline{3-6} 
\multicolumn{1}{|l|}{} &
   &
  \multicolumn{1}{c|}{\textbf{0 dB}} &
  \multicolumn{1}{c|}{\textbf{5 dB}} &
  \multicolumn{1}{c|}{\textbf{10 dB}} &
  \textbf{20 dB} \\ \hline
\multirow{4}{*}{\textbf{GER}} &
  \textbf{Noisy} &
  \multicolumn{1}{c|}{\begin{tabular}[c]{@{}c@{}}92.48\\ (89)\end{tabular}} &
  \multicolumn{1}{c|}{\begin{tabular}[c]{@{}c@{}}92.6\\ (89.22)\end{tabular}} &
  \multicolumn{1}{c|}{\begin{tabular}[c]{@{}c@{}}92.86\\ (89.78)\end{tabular}} &
  \begin{tabular}[c]{@{}c@{}}92.94\\ (90.02)\end{tabular} \\ \cline{2-6} 
 &
  \textbf{Diffusion} &
  \multicolumn{1}{c|}{\textbf{\begin{tabular}[c]{@{}c@{}}92.65\\ (89.3)\end{tabular}}} &
  \multicolumn{1}{c|}{\textbf{\begin{tabular}[c]{@{}c@{}}92.75\\ (89.5)\end{tabular}}} &
  \multicolumn{1}{c|}{\textbf{\begin{tabular}[c]{@{}c@{}}92.9\\ (89.9)\end{tabular}}} &
  \textbf{\begin{tabular}[c]{@{}c@{}}93.55\\ (90.5)\end{tabular}} \\ \cline{2-6} 
 &
  \textbf{Denoiser} &
  \multicolumn{1}{c|}{\begin{tabular}[c]{@{}c@{}}92.2\\ (88.8)\end{tabular}} &
  \multicolumn{1}{c|}{\begin{tabular}[c]{@{}c@{}}92.6\\ (89.25)\end{tabular}} &
  \multicolumn{1}{c|}{\begin{tabular}[c]{@{}c@{}}92.9\\ (89.5)\end{tabular}} &
  \begin{tabular}[c]{@{}c@{}}93.4\\ (89.95)\end{tabular} \\ \cline{2-6} 
 &
  \textbf{DTLN} &
  \multicolumn{1}{c|}{\begin{tabular}[c]{@{}c@{}}91.8\\ (88.2)\end{tabular}} &
  \multicolumn{1}{c|}{\begin{tabular}[c]{@{}c@{}}92.1\\ (88.7)\end{tabular}} &
  \multicolumn{1}{c|}{\begin{tabular}[c]{@{}c@{}}92.35\\ (89.3)\end{tabular}} &
  \begin{tabular}[c]{@{}c@{}}92.6\\ (89.7)\end{tabular} \\ \hline
\multirow{4}{*}{\textbf{ITA}} &
  \textbf{Noisy} &
  \multicolumn{1}{c|}{\begin{tabular}[c]{@{}c@{}}92.3\\ (88.92)\end{tabular}} &
  \multicolumn{1}{c|}{\begin{tabular}[c]{@{}c@{}}92.58\\ (89.22)\end{tabular}} &
  \multicolumn{1}{c|}{\begin{tabular}[c]{@{}c@{}}92.86\\ (89.24)\end{tabular}} &
  \begin{tabular}[c]{@{}c@{}}92.46\\ (89.4)\end{tabular} \\ \cline{2-6} 
 &
  \textbf{Diffusion} &
  \multicolumn{1}{c|}{\textbf{\begin{tabular}[c]{@{}c@{}}92.4\\ (88.95)\end{tabular}}} &
  \multicolumn{1}{c|}{\textbf{\begin{tabular}[c]{@{}c@{}}92.65\\ (89.25)\end{tabular}}} &
  \multicolumn{1}{c|}{\textbf{\begin{tabular}[c]{@{}c@{}}92.89\\ (89.5)\end{tabular}}} &
  \textbf{\begin{tabular}[c]{@{}c@{}}93\\ (89.55)\end{tabular}} \\ \cline{2-6} 
 &
  \textbf{Denoiser} &
  \multicolumn{1}{c|}{\begin{tabular}[c]{@{}c@{}}90.3\\ (85.75)\end{tabular}} &
  \multicolumn{1}{c|}{\begin{tabular}[c]{@{}c@{}}90.8\\ (86.35)\end{tabular}} &
  \multicolumn{1}{c|}{\begin{tabular}[c]{@{}c@{}}91.1\\ (87.2)\end{tabular}} &
  \begin{tabular}[c]{@{}c@{}}91.45\\ (87.5)\end{tabular} \\ \cline{2-6} 
 &
  \textbf{DTLN} &
  \multicolumn{1}{c|}{\begin{tabular}[c]{@{}c@{}}90.45\\ (85.8)\end{tabular}} &
  \multicolumn{1}{c|}{\begin{tabular}[c]{@{}c@{}}90.95\\ (86)\end{tabular}} &
  \multicolumn{1}{c|}{\begin{tabular}[c]{@{}c@{}}91.15\\ (86.8)\end{tabular}} &
  \begin{tabular}[c]{@{}c@{}}91.3\\ (87.1)\end{tabular} \\ \hline
\end{tabular}
\end{table}
\vspace{-0.35cm}
In this section, we first compare the performance of three different SE models (CDiffuSE, Denoiser, and DTLN) for the task of stress detection. The evaluations are carried out on both the GER and ITA datasets across four different signal-to-noise ratio (SNR) levels. Following the SE models comparison, we present and discuss the classification accuracies obtained, highlighting the impact of each feature set. Subsequently, we discuss the results of the perceptual study to verify if the patterns observed in the automatic stress detection models align with human perception. 

\vspace{-0.2cm}
\subsection{Comparison of different SE models:}
\vspace{-0.1cm}
This comparison is conducted using both sets of features
\subsubsection{\textbf{Heuristics-based features}}
Table \ref{tab:my-table1} presents the classification accuracies obtained from noisy and enhanced audio processed by different SE models, with and without post-processing (values in brackets). These accuracies were measured using a VAE+DNN classifier with acoustic and contextual features.
The results show that the addition of noise slightly affects stress detection performance in both the GER and ITA datasets. However, the impact of noise on the computation of acoustic features is not substantial, resulting in accuracies that are nearly on par with those obtained from clean audio. When using the Denoiser and DTLN SE models, there is a noticeable degradation in performance after enhancement, observed across all SNR levels. This degradation is especially evident in the ITA dataset, where the drop in accuracy is more significant. Interestingly, the CDiffuSE model exhibits a different trend. Despite the presence of noise, the CDiffuSE model not only preserves but actually improves stress detection performance compared to clean audio. This improvement may be attributed to the model's ability to effectively enhance the audio while retaining and possibly accentuating stress-related features, leading to superior classification outcomes. This suggests that the CDiffuSE model is particularly well-suited for this task, outperforming the other SE models in maintaining and enhancing stress properties in the audio. 

\subsubsection{\textbf{Self-supervised representations (wav2vec 2.0)}}
Table \ref{tab:my-table2} presents the classification results using wav2vec 2.0 features under both noisy and speech-enhanced conditions using VAE+DNN as classifier for stress detection task. The results show that the degradation in performance using wav2vec 2.0 features due to noise is significantly greater compared to the acoustic and context features. Although wav2vec 2.0 is generally reported in the literature to outperform acoustic and contextual features, the presence of noise severely impacts its feature extraction process, which in turn affects the stress detection accuracy. This indicates that wav2vec 2.0 features are more susceptible to noise and less robust in low SNR conditions. However, as the SNR increases, the performance of noisy data processed with wav2vec 2.0 features surpasses that of acoustic features. After applying speech enhancement, all SE models show an improvement over the noisy data. Among the SE models, the CDiffuSE model performs the best in 3 out of 4 scenarios for GER and in 2 out of 4 cases for ITA. This suggests that while wav2vec 2.0 features are sensitive to noise, their performance can be significantly enhanced with effective noise reduction techniques, particularly using the CDiffuSE model. Further, it is observed that the effect of postprocessing is more in 0 dB scenario in both GER and ITA which once again proves the drastic decline of the wav2vec 2.0 feature performance in low SNR scenarios in the stress detection task.
\begin{table}[]
\scriptsize
\centering
\captionsetup{labelfont=normalfont,textfont=normalfont}
\captionsetup{justification=centering}
\caption{\footnotesize{Classification accuracies with and without (in brackets) postprocessing, using wav2vec 2.0 features, for speech processed by different SE models under various SNR conditions. Accuracies for GER and ITA clean data are 94.38 (91.62) and 93.63 (91.23) respectively}}.
\begin{tabular}{|l|l|cccc|}
\hline
\multirow{2}{*}{Wave2vec 2.0} &
  \multirow{2}{*}{Condition} &
  \multicolumn{4}{c|}{SNR (dB)} \\ \cline{3-6} 
 &
   &
  \multicolumn{1}{l|}{0 dB} &
  \multicolumn{1}{l|}{5 dB} &
  \multicolumn{1}{l|}{10 dB} &
  \multicolumn{1}{l|}{20 dB} \\ \hline
\multirow{4}{*}{GER} &
  Noisy &
  \multicolumn{1}{c|}{\begin{tabular}[c]{@{}c@{}}87.12\\ (82.1)\end{tabular}} &
  \multicolumn{1}{c|}{\begin{tabular}[c]{@{}c@{}}91.52\\ (87.9)\end{tabular}} &
  \multicolumn{1}{c|}{\begin{tabular}[c]{@{}c@{}}93.72\\ (91.38)\end{tabular}} &
  \begin{tabular}[c]{@{}c@{}}94.34\\ (92.16)\end{tabular} \\ \cline{2-6} 
 &
  Diffusion &
  \multicolumn{1}{c|}{\begin{tabular}[c]{@{}c@{}}89.4\\ (85.2)\end{tabular}} &
  \multicolumn{1}{c|}{\textbf{\begin{tabular}[c]{@{}c@{}}93.2\\ (90.26)\end{tabular}}} &
  \multicolumn{1}{c|}{\textbf{\begin{tabular}[c]{@{}c@{}}94.23\\ (91.64)\end{tabular}}} &
  \textbf{\begin{tabular}[c]{@{}c@{}}94.8\\ (92.9)\end{tabular}} \\ \cline{2-6} 
 &
  Denoiser &
  \multicolumn{1}{c|}{\textbf{\begin{tabular}[c]{@{}c@{}}91.42\\ (87.46)\end{tabular}}} &
  \multicolumn{1}{c|}{\begin{tabular}[c]{@{}c@{}}92.9\\ (90.1)\end{tabular}} &
  \multicolumn{1}{c|}{\begin{tabular}[c]{@{}c@{}}94.1\\ (91.58)\end{tabular}} &
  \begin{tabular}[c]{@{}c@{}}94.3\\ (91.8)\end{tabular} \\ \cline{2-6} 
 &
  DTLN &
  \multicolumn{1}{c|}{\begin{tabular}[c]{@{}c@{}}89.18\\ (84.46)\end{tabular}} &
  \multicolumn{1}{c|}{\begin{tabular}[c]{@{}c@{}}92.5\\ (88.7)\end{tabular}} &
  \multicolumn{1}{c|}{\begin{tabular}[c]{@{}c@{}}93.5\\ (90.9)\end{tabular}} &
  \begin{tabular}[c]{@{}c@{}}94.1\\ (91.7)\end{tabular} \\ \hline
\multirow{4}{*}{ITA} &
  Noisy &
  \multicolumn{1}{c|}{\begin{tabular}[c]{@{}c@{}}85.96\\ (79.9)\end{tabular}} &
  \multicolumn{1}{c|}{\begin{tabular}[c]{@{}c@{}}91.4\\ (86.32)\end{tabular}} &
  \multicolumn{1}{c|}{\begin{tabular}[c]{@{}c@{}}93.3\\ (89.32)\end{tabular}} &
  \begin{tabular}[c]{@{}c@{}}94.28\\ (91.44)\end{tabular} \\ \cline{2-6} 
 &
  Diffusion &
  \multicolumn{1}{c|}{\begin{tabular}[c]{@{}c@{}}90.3\\ (86.8)\end{tabular}} &
  \multicolumn{1}{c|}{\begin{tabular}[c]{@{}c@{}}91.8\\ (88)\end{tabular}} &
  \multicolumn{1}{c|}{\textbf{\begin{tabular}[c]{@{}c@{}}93.4\\ (89.5)\end{tabular}}} &
  \textbf{\begin{tabular}[c]{@{}c@{}}94.5\\ (91.8)\end{tabular}} \\ \cline{2-6} 
 &
  Denoiser &
  \multicolumn{1}{c|}{\textbf{\begin{tabular}[c]{@{}c@{}}91.4\\ (86.64)\end{tabular}}} &
  \multicolumn{1}{c|}{\textbf{\begin{tabular}[c]{@{}c@{}}92.74\\ (88.76)\end{tabular}}} &
  \multicolumn{1}{c|}{\begin{tabular}[c]{@{}c@{}}92.9\\ (89.08)\end{tabular}} &
  \begin{tabular}[c]{@{}c@{}}94.22\\ (89.38)\end{tabular} \\ \cline{2-6} 
 &
  DTLN &
  \multicolumn{1}{c|}{\begin{tabular}[c]{@{}c@{}}86.48\\ (81.24)\end{tabular}} &
  \multicolumn{1}{c|}{\begin{tabular}[c]{@{}c@{}}91.34\\ (86.38)\end{tabular}} &
  \multicolumn{1}{c|}{\begin{tabular}[c]{@{}c@{}}92.6\\ (88.62)\end{tabular}} &
  \begin{tabular}[c]{@{}c@{}}93.72\\ (90.7)\end{tabular} \\ \hline
\end{tabular}
\end{table}

\vspace{-0.1cm}
\begin{table}[]
\scriptsize
\captionsetup{labelfont=normalfont,textfont=normalfont}
\captionsetup{justification=centering}
\caption{\footnotesize{Perceptual study based similarity statistics of speech processed SE models (CDiffuSE, Denoiser, and DTLN) with clean speech}}.
\centering
\begin{tabular}{|l|cc|cc|}
\hline
\multicolumn{1}{|c|}{\multirow{2}{*}{}} & \multicolumn{2}{c|}{\textbf{ITA}}  & \multicolumn{2}{c|}{\textbf{GER}}  \\ \cline{2-5} 
\multicolumn{1}{|c|}{} & \multicolumn{1}{c|}{\textbf{Perceptual Study}} & \textbf{Accuracy} & \multicolumn{1}{c|}{\textbf{Perceptual Study}} & \textbf{Accuracy} \\ \hline
\textbf{Diffusion}                      & \multicolumn{1}{c|}{45.91\%} & 82.65\% & \multicolumn{1}{c|}{37.60\%} & 76.86\% \\ \hline
\textbf{Denoiser}                       & \multicolumn{1}{c|}{30.34\%} & 66.92\% & \multicolumn{1}{c|}{34.88\%} & 72.62\% \\ \hline
\textbf{DTLN}                           & \multicolumn{1}{c|}{23.75\%} & 63.88\% & \multicolumn{1}{c|}{27.52\%} & 67.95\% \\ \hline
\end{tabular}
\end{table}



\subsection{Perceptual study results:}
\vspace{-0.1cm}
Table III presents the results of both the perceptual study and the automatic syllable stress detection performance on the sub-set used in the study for three different speech enhancement (SE) models. The perceptual study results show the percentage of enhanced audio samples rated as most similar to the clean reference audio, while the accuracy values represent the performance of the VAE+DNN model on the same set of enhanced audio samples. The perceptual study reveals that CDiffuSE-enhanced audio achieved the highest similarity to clean audio, with DTLN showing the lowest similarity. This trend is also reflected in the classification accuracies from the stress detection task. For the ITA dataset, CDiffuSE-enhanced audio reached an accuracy of 82.65\%, outperforming Denoiser (66.92\%) and DTLN (63.88\%). In the GER dataset, CDiffuSE also led with an accuracy of 76.86\%, followed by Denoiser at 72.62\% and DTLN at 67.95\%. The lower accuracy in the GER dataset suggests that SE models negatively affect stress detection more in this dataset compared to ITA, highlighting the difficulties in enhancing speech for stress detection under noisy conditions for certain languages.



\vspace{-0.2cm}
\section{Conclusion}
\vspace{-0.1cm}
In this study, we evaluated the impact of three different SE models (DTLN, Denoiser, and CDiffuSE) in preserving the stress cues for the task of automatic syllable stress detection under noisy conditions (gaussian noise with different SNRs). We also conducted a perceptual study to determine which SE model produces enhanced audio without introducing artifacts that could alter the word's meaning (which can belong to multiple grammatical categories based on stress placement) and remains similar to the clean audio. Experiments are performed on ISLE corpus of non-native speakers (German, and Italian) speaking English. We considered two sets of features, 1) heuristic-based features, and 2) self-supervised representations (wav2vec 2.0). From the experimental results, stress detection performance remains robust when heuristic features are used compared to self-supervised features. Further, it is observed that audios enhanced with the SE model belonging to generative modeling category (CDiffuSE) outperforms the discriminative SE models (DTLN and Denoiser) for stress detection task. And this pattern of performance is similar in perceptual study outcomes as well.


\bibliographystyle{IEEEbib}
\bibliography{main}

\end{document}